\documentclass[preprint,authoryear,12pt]{elsarticle}
\journal{International Journal of Solids and Structures}
\usepackage{graphics,epsfig}
\usepackage{graphicx}
\usepackage{amssymb,amstext,amsmath}
\usepackage{booktabs}
\usepackage{natbib}
\usepackage[latin1]{inputenc}
\usepackage{epstopdf}

\begin{document}
\begin{frontmatter}
\title{A single crystal beam bent in double slip}
\author{X. Cui, K. C. Le\footnote{corresponding author: ++49 234 32-26033, email: chau.le@rub.de}}
\address{Lehrstuhl f\"{u}r Mechanik - Materialtheorie, Ruhr-Universit\"{a}t Bochum,\\D-44780 Bochum, Germany}
\begin{abstract} 
The theory of plastic bending of single crystal beam having two active slip systems is proposed. Applying the variational-asymptotic method we reduce the energy functional of the beam to the one-dimensional energy functional which admits analytical solutions for the symmetric slip systems. The threshold value at the onset of plastic yielding, the dislocation density, as well as the moment-curvature curve are found. We solve also a similar problem that takes the energy dissipation into account.
\end{abstract}
\end{frontmatter}
\section{Introduction}
When a single crystal beam is plastically bent, excess dislocations occur on certain active slip planes to reduce its energy. As a rule, the number of dislocations is huge, so, instead of a difficult task of describing individual dislocations, the use of continuum approach to find the dislocation densities in the bent state turns out more effective. The first attempt of taking into account continuously distributed excess dislocations in the plastically bent beam was made by \citet{Nye1953} who expressed the curvature of the beam caused by dislocations in terms of the dislocation density tensor bearing now his name. \citet{Read1957} and \citet{Bilby1958} have extended this result to the case when the stress due to excess dislocations does not vanish (see also \citep{Ashby1970,Evans1995,Mura1987,Wang2003}). Unfortunately, because of the lack of physically sound continuum dislocation theories, no quantitative dislocation distribution could be predicted in terms of the bending moment. The first quantitative prediction of the dislocation density of single crystal beam bent in single slip has been made in \citep{Le-nguyen2013}. The analyzed model was based on the continuum dislocation theory proposed by \citet{Berdichevsky2006a, Berdichevsky2006b} and developed further in \citep{Berdichevsky-Le2007,Kaluza-Le2011,Kochmann-le2008,Kochmann-le2009a,Kochmann-le2009b,Le-sembiring2008a,Le-sembiring2008b, Le-sembiring2009} (see also the alternative approaches in \citep{Acharya2000,Gurtin2002,Gurtin2007}). The information about dislocation distribution obtained in \citep{Le-nguyen2013} has been used to explain the formation of small angle tilt boundaries and predict the number of polygons in a polygonized beam after annealing (see also \citep{Le-nguyen2011,Le-nguyen2012}).

The real bending of single crystal beam is often more involved because two or more slip systems could be activated during the plastic bending, depending on the orientation of crystal with respect to the beam axis affecting the Schmid factors. For instance, if the fcc single crystal beam having the axis parallel to $[\bar{1}10]$-direction is bent about the $[110]$ direction, then there are two symmetric active slip systems oriented at angles $\pm 54.7^\circ $ with respect to the beam axis (see, e.g.,  \citep{Kysar2010}). Simulations of such bending problems require more elaborate models because of the interaction between edge dislocations of different groups. This paper aims at  constructing an approximate one-dimensional theory of bending of single crystal beams having two active slip systems and, thus, extending the results obtained in \citep{Le-nguyen2013} to double slip. We incorporate the interactions of excess dislocations in a material model similar to that proposed by \citet{Le-sembiring2009}. We then use the variational-asymptotic method developed in \citep{Berdichevsky1983,Le1999} to derive the one-dimensional approximate theories from the exact three-dimensional problems in two cases: (i) the resistance to dislocation motion is negligibly small so that the displacements and plastic slips can be found by minimizing the energy functional, (ii) the rate-independent dissipation is taken into account. The obtained one-dimensional variational problems admit analytical solutions representing the smooth minimizers for the symmetric active slip systems. It is established that there exist energetic as well as dissipative thresholds for the dislocation nucleation which depend on the thickness of the beam. This exhibits the typical size effect of the gradient theory. Based on this analytical solution the deflection of the beam, the dislocation density, and the moment-curvature curve are analyzed in terms of the bending moment for different loading/unloading processes.

The paper is organized as follows. In the next Section the 3-D models of single crystal beam with continuously distributed dislocations bent in double slip are formulated. Sections 3 and 4 derive the one-dimensional equations and solve them for the case of zero dissipation. Sections 5 and 6 consider the same problems but with dissipation. Finally, Section 7 concludes the paper. 

\section{3-D models of crystal beam bent in double slip}

\begin{figure}[htb]
    \begin{center}
    \includegraphics[width=11cm]{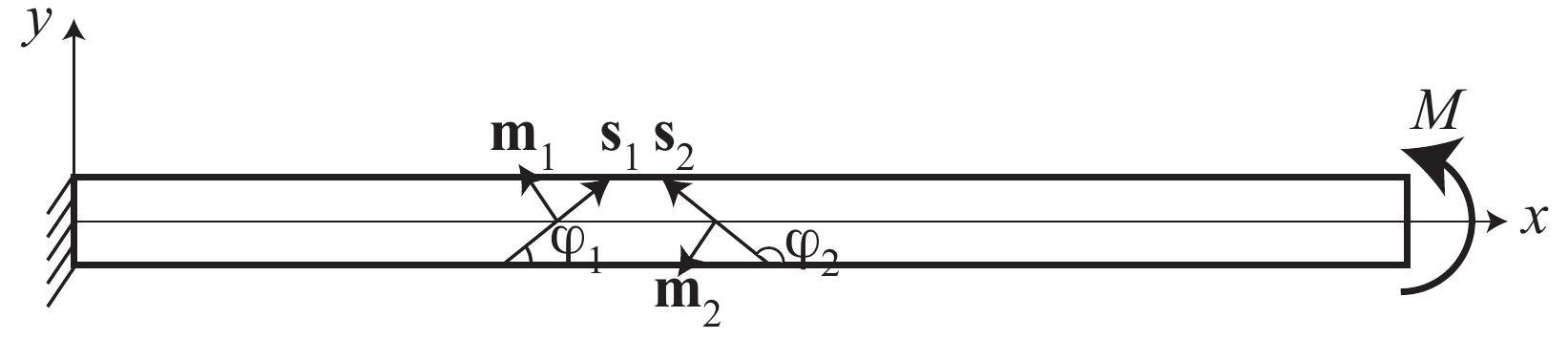}
    \end{center}
    \caption{Single crystal beam bent in double slip by a moment $M$.}
    \label{fig:1}
\end{figure}

Consider a single crystal beam bent by a moment $M$. In the undeformed state the beam occupies the domain $\mathcal{V}$ of the three-dimensional euclidean space, with the cartesian coordinates $(x,y,z)\in (0,L)\times (-h/2,h/2)\times (0,a)$ where $L$, $h$, and $a$ are the length, height, and width of the beam, respectively. We will assume that $h\ll a \ll L$ and that the beam is bent under the plane strain condition. For definiteness, let the beam be clamped at $x = 0$ and subjected to a linearly distributed traction $-\tau y$ having the resultant moment $M$ at $x = L$. Besides, the upper and lower boundaries of the beam at $y= h/2$ and $y= - h/2$ are traction free. If the applied moment is small, it is natural to expect that the beam deforms elastically and the stress distribution is linear over the thickness according to the elementary beam theory. However, if the applied moment exceeds some critical value, edge dislocations with dislocation lines parallel to the $z$-axis may appear to reduce energy of the bent beam. We assume that, at the onset of yielding and during the plastic deformations, the crystal admits two active slip systems with slip planes inclined at angles $\varphi _1$ and $\varphi _2$ to the plane $y=0$ as shown in Fig.~\ref{fig:1}. 

Under the plane strain state condition there are only two non-zero components of the displacements that do not depend on $z$, $u_x(x,y)$ and $u_y(x,y)$. Consequently, the non-zero components of the total strain tensor are 
\begin{equation*}
\varepsilon _{xx}=u_{x,x},\quad \varepsilon _{yy}=u_{y,y},\quad \varepsilon _{xy}=\varepsilon _{yx}=\frac{1}{2}(u_{x,y}+u_{y,x}).
\end{equation*}
Throughout the paper the comma before an index is used to denote the partial derivative with respect to the corresponding coordinate. Since there are two active slip systems, the plastic distortion tensor is given by
\begin{equation*}
\beta _{ij}=\beta _1(x,y)s^1_im^1_j+\beta _2(x,y)s^{2}_{i}m^{2}_{j},
\end{equation*}
with $\mathbf{s}^{1} = (\cos \varphi_{1},\sin \varphi_{1},0)$, $\mathbf{m}^{1} = (-\sin \varphi_{1},\cos \varphi_{1},0)$, $\mathbf{s}^{2} = (\cos \varphi_{2},\sin \varphi_{2},0)$, $\mathbf{m}^{2} = (-\sin \varphi_{2},\cos \varphi_{2},0)$ being the unit vectors denoting the slip direction and the normal to the slip planes of the corresponding slip system. We call $\beta _1(x,y)$ and $\beta _2(x,y)$ the plastic slips. Thus, the finding of plastic distortion tensor reduces to determining the functions $\beta _1(x,y)$ and $\beta _2(x,y)$. Non-zero components of the plastic strain tensor, $\varepsilon ^p_{ij}=\frac{1}{2}(\beta _{ij}+\beta _{ji})$, read
\begin{align*}&\varepsilon^{p}_{xx} = -\frac{1}{2}\beta_{1}\sin 2\varphi_{1}-\frac{1}{2}\beta_{2}\sin 2\varphi_{2},\quad \varepsilon^{p}_{yy} = \frac{1}{2}\beta_{1}\sin 2\varphi_{1}+\frac{1}{2}\beta_{2}\sin 2\varphi_{2},\\
&\varepsilon^{p}_{xy} = \varepsilon^{p}_{yx} = \frac{1}{2}\beta_{1}\cos 2\varphi_{1}+\frac{1}{2} \beta_{2}\cos 2\varphi_{2}.
\end{align*}
Accordingly, the non-zero components of the elastic strain tensor, $\varepsilon ^{e}_{ij}=\varepsilon _{ij}-\varepsilon ^{p}_{ij}$, are
\begin{align*}
&\varepsilon^{e}_{xx} = u_{x,x}+\frac{1}{2}\beta_{1}\sin 2\varphi_{1}+\frac{1}{2}\beta_{2}\sin 2\varphi_{2}, 
\\
&\varepsilon^{e}_{yy} = u_{y,y}-\frac{1}{2}\beta_{1}\sin 2\varphi_{1}-\frac{1}{2}\beta_{2}\sin 2\varphi_{2},\\
&\varepsilon^{e}_{xy} = \varepsilon^{e}_{yx} = \frac{1}{2}\left( u_{x,y}+u_{y,x}-\beta_{1}\cos 2\varphi_{1}-\beta_{2}\cos 2\varphi_{2} \right) .
\end{align*}

The distribution of excess dislocations associated with these two active slip systems is described by the dislocation density tensor (introduced by \citet{Nye1953}, \citet{Bilby1954}, and \citet{Kroener1955}), $\alpha _{ij}=\varepsilon _{jkl}\beta _{il,k}$, whose non-zero components read
\begin{align*}
&\alpha_{xz} = \beta _{1,x}\cos^{2}\varphi_{1}+\beta_{1,y}\cos \varphi_{1}\sin \varphi_{1}+\beta_{2,x}\cos^{2}\varphi_{2}+\beta_{2,y}\cos \varphi_{2}\sin \varphi_{2} ,
\\
&\alpha_{yz} =  \beta_{1,x}\cos \varphi_{1} \sin \varphi_{1}+\beta_{1,y} \sin^{2} \varphi_{1}+\beta_{2,x}\cos \varphi_{2}\sin \varphi_{2}+\beta_{2,y}\sin^{2} \varphi_{2}.
\end{align*}
Thus, we see that there are two different groups of excess edge dislocations associated with the slip systems 1 and 2, whose resultant Burgers' vectors show in the directions $\mathbf{s}^1$ and $\mathbf{s}^2$, respectively. These give two scalar dislocation densities
\begin{align}\label{2.1}
\rho_{1} = \frac{1}{b} \vert \beta_{1,x}\cos \varphi_{1}+\beta_{1,y}\sin \varphi_{1} \vert, \quad\rho_{2} = \frac{1}{b}\vert \beta_{2,x}\cos \varphi_{2}+\beta_{2,y}\sin \varphi_{2} \vert ,
\end{align}
with $b$ being the magnitude of the Burgers vector.

Let us propose the free energy density of the single crystal deforming in double slip in the form (see \citep{Le-sembiring2009})
\begin{equation}\label{2.2}
\phi = \frac{1}{2} \lambda (\varepsilon^{e}_{kk})^{2}+\mu \varepsilon^{e}_{ij} \varepsilon^{e}_{ij}+\mu k \ln\frac{1}{1-\frac{\rho_{1}}{\rho_{s}}}+\mu k \ln \frac{1}{1-\frac{\rho_{2}}{\rho_{s}}}+\mu \chi\frac{\rho_{1}\rho_{2}}{\rho^{2}_{s}},
\end{equation}
with $\lambda$ and $\mu$ being the Lam\'e constants, $\rho_{s}$ the saturated dislocation density, and $k$ and $\chi $ the constants of the material. The first two terms in \eqref{2.2} represent the contribution to the energy of crystal by the elastic strain. The third and fourth terms are the contributions to the energy from the two separate groups of excess dislocations \citep{Berdichevsky2006b}. Finally, the last term takes the interaction of dislocations between these two groups into account. For small up to moderate dislocation densities the logarithmic terms can be approximated by the Taylor series as follows
\[\ln \frac{1}{1-\frac{\rho}{\rho_{s}}} \cong \frac{\rho}{\rho_{s}}+\frac{1}{2}\dfrac{\rho^{2}}{\rho^{2}_{s}}.\]
We shall use further only this approximation.

With \eqref{2.1} and \eqref{2.2} the total energy functional of the bent beam becomes
\begin{multline}
I = a\int^{L}_{0}\int^{h/2}_{-h/2} \bigg[ \frac{1}{2} \lambda (u_{x,x}+u_{y,y}) ^{2} +\mu ( u_{x,x}+\frac{1}{2}\beta_{1}\sin 2\varphi_{1}+\frac{1}{2}\beta_{2}\sin 2\varphi_{2}) ^{2}   
\\
+\mu (u_{y,y}-\frac{1}{2}\beta_{1}\sin 2\varphi_{1}-\frac{1}{2}\beta_{2}\sin 2\varphi_{2}) ^{2}+\frac{1}{2}\mu ( u_{x,y}+u_{y,x}-\beta_{1}\cos 2\varphi_{1}
\\
-\beta_{2}\cos 2\varphi_{2})^{2}+\mu k\frac{\vert\beta_{1,x}\cos \varphi_{1}+\beta_{1,y}\sin\varphi_{1}\vert}{b\rho_{s}}+\frac{1}{2}\mu k \frac{\left( \beta_{1,x}\cos \varphi_{1}+\beta_{1,y}\sin\varphi_{1}\right)^{2} }{b^{2}\rho^{2}_{s}}
\\
+\mu k \frac{\vert\beta_{2,x}\cos\varphi_{2}+\beta_{2,y}\sin \varphi_{2} \vert}{b\rho_{s}}+\frac{1}{2}\mu k\frac{\left( \beta_{2,x}\cos\varphi_{2}+\beta_{2,y}\sin\varphi_{2}\right)^{2} }{b^{2}\rho^{2}_{s}}
\\
+k\mu\chi\frac{\vert\beta_{1,x}\cos\varphi_{1}+\beta_{1,y}\sin\varphi_{1}\vert \vert\beta_{2,x}\cos \varphi_{2}+\beta_{2,y}\sin \varphi_{2}\vert}{b^{2}\rho^{2}_{s}} \bigg] dxdy 
\\
+a\int^{h/2}_{-h/2}\tau yu_x |_{x=L}dy. \label{2.3}
\end{multline}
The last term in \eqref{2.3} is the work done by the linearly distributed traction $-\tau y$ acting at the boundary $x=L$.  If the dissipation caused by the dislocation motion is negligible, then the true displacements $u_x$, $u_y$ and plastic slips $\beta_1$, $\beta _2$ in the {\it final} state of bending minimize energy functional \eqref{2.3} among all admissible displacements and plastic slips satisfying the kinematic boundary condition. By the final state of bending we means the equilibrium state which is established at fixed loading condition after dislocation nucleation and after the movement of dislocations toward their equilibrium positions is finished. Thus, the whole process of dislocation nucleation and dislocation dynamics cannot be described by this theory. The bending moment $M=\tau ah^3/12$ is regarded as a control parameter, so one can study the evolution of the dislocation network in terms of  $M$.

If the dissipation due to dislocation motion cannot be neglected, energy minimization should be replaced by the variational equation \citep{Sedov1965}
\begin{align}\label{2.4}
\delta I+a\int^{L}_{0}\int^{h/2}_{-h/2}(\frac{\partial D}{\partial\dot{\beta _1}}\delta\beta _1+\frac{\partial D}{\partial\dot{\beta _2}} \delta \beta _2)dxdy = 0.
\end{align}
The last integral in this equation describes the energy dissipation due to dislocation motion, where the dissipation function $D(\dot{\beta }_1,\dot{\beta }_2)$ is assumed to depend only on the plastic slip rates. We shall consider the simplest rate-independent theory for which
\begin{align*}
D(\dot{\beta }_1,\dot{\beta }_2) = K(\vert \dot{\beta}_{1}\vert+\vert \dot{\beta}_{2}\vert),
\end{align*}
with $K$ being the critical resolved shear stress. Then, provided the signs of $\dot{\beta }_1$ and $\dot{\beta }_2$ do not change during the evolution of $\beta _1$ and $\beta _2$, the variational equation \eqref{2.4} reduces to minimizing the following ``relaxed energy'' functional
\begin{multline}
I_d = a\int^{L}_{0}\int^{h/2}_{-h/2} \bigg[ \frac{1}{2} \lambda (u_{x,x}+u_{y,y}) ^{2} +\mu ( u_{x,x}+\frac{1}{2}\beta_{1}\sin 2\varphi_{1}+\frac{1}{2}\beta_{2}\sin 2\varphi_{2}) ^{2}   
\\
+\mu (u_{y,y}-\frac{1}{2}\beta_{1}\sin 2\varphi_{1}-\frac{1}{2}\beta_{2}\sin 2\varphi_{2}) ^{2}+\frac{1}{2}\mu ( u_{x,y}+u_{y,x}-\beta_{1}\cos 2\varphi_{1}
\\
-\beta_{2}\cos 2\varphi_{2})^{2}+\mu k\frac{\vert\beta_{1,x}\cos \varphi_{1}+\beta_{1,y}\sin\varphi_{1}\vert}{b\rho_{s}}+\frac{1}{2}\mu k \frac{\left( \beta_{1,x}\cos \varphi_{1}+\beta_{1,y}\sin\varphi_{1}\right)^{2} }{b^{2}\rho^{2}_{s}}
\\
+\mu k \frac{\vert\beta_{2,x}\cos\varphi_{2}+\beta_{2,y}\sin \varphi_{2} \vert}{b\rho_{s}}+\frac{1}{2}\mu k\frac{\left( \beta_{2,x}\cos\varphi_{2}+\beta_{2,y}\sin\varphi_{2}\right)^{2} }{b^{2}\rho^{2}_{s}}
\\
+k\mu\chi\frac{\vert\beta_{1,x}\cos\varphi_{1}+\beta_{1,y}\sin\varphi_{1}\vert \vert\beta_{2,x}\cos \varphi_{2}+\beta_{2,y}\sin \varphi_{2}\vert}{b^{2}\rho^{2}_{s}}  
\\
+K\, \text{sign}(\dot{\beta}_{1}) \beta_{1}+K\, \text{sign}(\dot{\beta}_{2}) \beta_{2} \bigg] dxdy+a\int^{h/2}_{-h/2}\tau yu_x |_{x=L}dy. \label{2.5}
\end{multline}
So, the true displacement and plastic slip fields in the {\it final} equilibrium state of plastic bending minimize the ``relaxed'' energy functional $I_d$ among all admissible displacements and plastic slips. Finally, if $\dot{\beta}_1=0$ and $\dot{\beta}_2=0$, then the plastic slips are frozen, while the displacements should be found by minimizing \eqref{2.3} with the frozen plastic slips. Let us mention here also the more complicate model taking into account the interaction of dislocation in dissipation leading to the latent hardening. In such model we assume the rate-idependent dissipation function in the form
\begin{equation*}
D(\dot{\beta }_1,\dot{\beta }_2) = K(\vert \dot{\beta}_{1}\vert+\vert \dot{\beta}_{2}\vert +\xi \sqrt{\vert \dot{\beta}_{1}\vert \vert \dot{\beta}_{2}\vert}).
\end{equation*}
In this paper we shall not dwell on this model.

\section{Energy minimization}
We first focus on the minimization problem \eqref{2.3}. It is convenient to introduce the following dimensionless variables and quantities 
\begin{align}\label{3.1}
&\bar{x}=b\rho_sx,\quad \bar{y}=b\rho_sy,\quad \bar{h}=b\rho_sh,\quad \bar{L}=b\rho_sL,\quad\bar{\tau}=\frac{\tau}{\mu b \rho_{s}},\nonumber
\\
&\bar{u}_{x} = b\rho_{s}u_{x},\quad \bar{u}_{y} = b\rho_{s}u_{y},\quad \bar{I} = \frac{I(b\rho_{s})^{2}}{\mu a},\quad \gamma = \frac{\lambda}{\mu},
\end{align} 
which simplify the minimization problem. Now the energy functional can be rewritten in the dimensionless form as follows
\begin{multline}\label{3.2}
I = \int^{L}_{0}\int^{h/2}_{-h/2} \bigg[ \frac{1}{2}\gamma (u_{x,x}+u_{y,y}) ^{2} +(u_{x,x}+\frac{1}{2} \beta_{1} \sin 2\varphi_{1}+\frac{1}{2}\beta_{2}\sin 2\varphi_{2}) ^{2} 
\\+(u_{y,y}-\frac{1}{2}\beta_{1}\sin 2\varphi_{1}-\frac{1}{2}\beta_{2}\sin 2\varphi_{2}) ^{2}+\frac{1}{2}( u_{x,y}+u_{y,x}-\beta_{1}\cos 2\varphi_{1}
\\-\beta_{2}\cos 2\varphi_{2})^{2}+k\vert\beta_{1,x}\cos \varphi_{1}+\beta_{1,y}\sin\varphi_{1}\vert+\frac{1}{2}k( \beta_{1,x}\cos \varphi_{1}+\beta_{1,y}\sin \varphi_{1})^{2}
\\+k\vert\beta_{2,x}\cos\varphi_{2}+\beta_{2,y}\sin\varphi_{2}\vert+\frac{1}{2}k( \beta_{2,x}\cos\varphi_{2}+\beta_{2,y}\sin\varphi_{2})^{2}+k\chi \vert \beta_{1,x}\cos\varphi_{1}
\\+\beta_{1,y}\sin\varphi_{1}\vert \vert\beta_{2,x}\cos\varphi_{2}+\beta_{2,y}\sin \varphi_{2}\vert \bigg]dxdy+\int^{h/2}_{-h/2}\tau yu_x |_{x=L}dy.
\end{multline}
The bar over the dimensionless quantities is dropped for short because in the subsequent analysis we shall deal only with them.

Energy functional \eqref{3.2} contains a small parameter $h$, so it can be reduced to 1-D energy functional by the variational asymptotic method (see \citep{Berdichevsky1983,Le1999}). For this purpose let us introduce the rescaled coordinate $\zeta =y/h$, $\zeta \in (-1/2,1/2)$ to fix the domain over the thickness in the passage to the limit $h\to 0$. With this new variable $\zeta $ the small parameter $h$ enters the functional explicitly through the formulas
\begin{align*}
u_{i,y} =  \frac{1}{h}u_{i,\zeta},\quad \beta_{\alpha ,y} = \frac{1}{h}\beta_{\alpha ,\zeta}, \quad \alpha =1,2.
\end{align*}
Since the boundary condition at $x = L$ does not influence the inner asymptotic distributions of the displacements and the plastic slips over the thickness, we put $ \tau = 0$ for a while. At the first step of the variational-asymptotic procedure we keep the asymptotic principal terms in \eqref{3.2} to obtain
\begin{multline*}
I_{0} = h\int^{L}_{0}\int^{1/2}_{1/2}\bigg[\frac{1}{2h^{2}}\gamma(u_{y,\zeta})^{2}+\frac{1}{h^{2}}(u_{y,\zeta})^{2}+\frac{1}{2h^{2}}(u_{x,\zeta})^{2}+\frac{k}{h}\vert\beta_{1,\zeta}\sin\varphi_{1}\vert
\\+\frac{k}{2h^{2}}(\beta_{1,\zeta}\sin\varphi_{1})^{2}+\frac{k}{h}\vert\beta_{2,\zeta}\sin\varphi_{2}\vert+\frac{k}{2h^{2}}(\beta_{2,\zeta}\sin\varphi_{2})^{2}
\\
+\frac{\chi k}{h^{2}}\vert\beta_{1,\zeta}\sin\varphi_{1}\vert
\vert\beta_{2,\zeta}\sin\varphi_{2}\vert\bigg]dxd\zeta .
\end{multline*}
Functional $I_{0}$ is positive definite, so its minimum must be zero and is achieved at
$u_{x,\zeta} = u_{y,\zeta} = \beta_{1,\zeta} = \beta_{2,\zeta} = 0.$ For the bending states which we are interested in let us set at this step $u_x = 0$ and $\beta _1=\beta _2=0$ to obtain
\begin{align*}
u_{x} = 0,\quad u_{y} = v(x),\quad \beta_{1} = \beta_{2} = 0.
\end{align*}
At the second step, we fix $v(x)$ and seek the minimizer in the form
\begin{align*}
u_{x} = u_{x}^\prime (x,\zeta),\quad u_{y} = v(x)+u_{y}^\prime (x,\zeta),\quad \beta_{1} = \beta_{1}^\prime (x,\zeta),\quad \beta_{2} = \beta_{2}^\prime (x,\zeta),
\end{align*}
with $ u_{x}^\prime (x,\zeta),u_{y}^\prime (x,\zeta),\beta_{1}^\prime (x,\zeta),\beta_{2}^\prime (x,\zeta) $ being the correction terms which are small compared to $v(x)$. Substituting these formulas into \eqref{3.2} and then keeping the asymptotically leading terms containing $u_x^\prime (x,\zeta )$, $u_y^\prime (x,\zeta )$, and $\beta _\alpha ^\prime (x,\zeta )$, we obtain
\begin{align*}
I_{1} = &h\int^{L}_{0}\int^{1/2}_{1/2}\bigg[\frac{1}{2h^{2}}\gamma(u_{y,\zeta}^\prime )^{2}+\frac{1}{h^{2}}(u_{y,\zeta}^\prime )^{2}+\frac{1}{2}( \frac{1}{h}u^\prime _{x,\zeta}+v_{,x})^{2}+\frac{k}{h}\vert\beta^\prime _{1,\zeta}\sin\varphi_{1}\vert\nonumber
\\&+\frac{k}{2h^{2}}(\beta^\prime _{1,\zeta}\sin\varphi_{1})^{2}+\frac{k}{h}\vert\beta^\prime _{2,\zeta}\sin\varphi_{2}\vert+\frac{k}{2h^{2}}(\beta^\prime _{2,\zeta}\sin\varphi_{2})^{2}+\frac{\chi k}{h^{2}}\vert\beta^\prime _{1,\zeta}\sin\varphi_{1}\vert\nonumber
\\& \vert\beta^\prime _{2,\zeta}\sin\varphi_{2}\vert\bigg]dxd\zeta
\end{align*}
Since functional $ I_{1} $ is also positive definite, its minimum is again zero and is achieved at
\begin{align*}
u_{y,\zeta}^\prime = 0,\quad u_{x,\zeta}^\prime = -hv_{,x},\quad \beta^\prime_{1,\zeta} = 0,\quad\beta^\prime_{2,\zeta} = 0.
\end{align*}
This implies
\begin{align*}
u_{y}^\prime = 0,\quad u_{x}^\prime = -h\zeta v_{,x},\quad \beta^\prime_{1} = 0,\quad\beta^\prime_{2} = 0.
\end{align*}
At the third step of the variational-asymptotic procedure, we look for the minimizer in the form
\begin{align}\label{3.3}
&u_{x} =  -h\zeta v_{,x}+u_{x}^{\prime \prime}(x,\zeta),\quad u_{y} = v(x)+u_{y}^{\prime \prime}(x,\zeta),
\\
&\beta_{1} = \beta_{1}^{\prime \prime}(x,\zeta),\quad\beta_{2} = \beta_{2}^{\prime \prime}(x,\zeta), \nonumber
\end{align}
with $ u_{x}^{\prime \prime}(x,\zeta)$, $u_{y}^{\prime \prime}(x,\zeta)$, $\beta_{1}^{\prime \prime}(x,\zeta) $, and $ \beta_{2}^{\prime \prime}(x,\zeta) $ being the correction terms which are assumed small compared with those found at the second step. By redefining $v(x)$ if required, we may put the following constraints on these correction terms
\begin{align}\label{3.4}
\langle u_{y}^{\prime \prime}\rangle = 0,\quad \langle \beta^{\prime \prime}_{1}\rangle = 0,\quad\langle \beta^{\prime \prime}_{2}\rangle = 0,\quad \mathrm{where}\quad \langle \cdot\rangle = \int^{1/2}_{-1/2}\cdot \, d\zeta .
\end{align}
Constraint $\langle u_y^{\prime \prime }\rangle =0$ means that $v(x)$ is the average deflection (over the thickness) of the beam. Substituting \eqref{3.3} into \eqref{3.2}, then keeping the asymptotically leading terms containing $u_x^{\prime \prime } (x,\zeta )$, $u_y^{\prime \prime } (x,\zeta )$, $\beta ^{\prime \prime }_1 (x,\zeta )$, and $\beta ^{\prime \prime }_2 (x,\zeta )$ to get
\begin{multline*}
I_{2} = h\int^{L}_{0}\int^{1/2}_{1/2}\bigg[\frac{1}{2}\gamma(-h\zeta v_{,xx}+\frac{1}{h}u_{y,\zeta}^{\prime \prime})^{2}+(-h\zeta v_{,xx}+\frac{\beta^{\prime \prime}_{1}}{2}\sin2\varphi_{1}+\frac{\beta^{\prime \prime}_{2}}{2}\sin 2\varphi_{2})^{2}
\\+( \frac{1}{h}u^{\prime \prime}_{y,\zeta}-\frac{1}{2}\beta^{\prime \prime}_{1}\sin 2\varphi_{1}-\frac{1}{2}\beta^{\prime \prime}_{2}\sin2\varphi_{2})^{2}+\frac{1}{2}(\frac{1}{h}u^{\prime \prime}_{x,\zeta}-\beta^{\prime \prime}_{1}\cos 2\varphi_{1}-\beta^{\prime \prime}_{2}\cos 2\varphi_{2})^2 
\\+\frac{k}{h}\vert\beta^{\prime \prime}_{1,\zeta}\sin\varphi_{1}\vert+\frac{k}{2h^{2}}(\beta^{\prime \prime}_{1,\zeta}\sin\varphi_{1})^{2}+\frac{k}{h}\vert\beta^{\prime \prime}_{2,\zeta}\sin\varphi_{2}\vert+\frac{k}{2h^{2}}(\beta^{\prime \prime}_{2,\zeta}\sin\varphi_{2})^{2}
\\+\frac{\chi k}{h^{2}}\vert\beta^{\prime \prime}_{1,\zeta}\sin\varphi_{1}\vert \vert\beta^{\prime \prime}_{2,\zeta}\sin\varphi_{2}\vert\bigg] dxd\zeta .
\end{multline*}
Functional $I_2$ can be reduced to a functional depending only on $\beta ^{\prime \prime }_1$ and $\beta ^{\prime \prime }_2$. Indeed, fixing first $\beta ^{\prime \prime }_1$ and $\beta ^{\prime \prime }_2$ and varying this functional with respect to $u_x^{\prime \prime }$ and $u_y^{\prime \prime }$, then using the natural boundary conditions at $\zeta =\pm 1/2$, we obtain,
\begin{equation}\label{3.5}
\begin{split}
\frac{1}{h}(\gamma+2)u^{\prime \prime }_{y,\zeta} = \gamma h\zeta v_{,xx}+\beta^{\prime \prime }_{1}\sin2\varphi_{1}+\beta^{\prime \prime }_{2}\sin2\varphi_{2},
\\
\frac{1}{h}u^{\prime \prime }_{x,\zeta} = \beta^{\prime \prime }_{1}\cos2\varphi_{1}+\beta^{\prime \prime }_{2}\cos2\varphi_{2}.
\end{split}
\end{equation}
After finding $u_x^{\prime \prime }$ and $u_y^{\prime \prime }$ according to these equations, we substitute \eqref{3.5} into \eqref{3.2} and change $\zeta$ back to $y$. Since the functional does not contain $\beta ^{\prime \prime }_{1,x}$ and $\beta ^{\prime \prime }_{2,x}$, the thickness problem reduces to minimizing the following functional with respect to $\beta ^{\prime \prime }_1(x,y)$ and $\beta ^{\prime \prime }_2(x,y)$
\begin{align*}
I_{3} &= \int_{-h/2}^{h/2}\bigg[ \frac{\kappa}{2}( -2v_{,xx}y+\beta^{\prime \prime}_{1}\sin2\varphi_{1}+\beta^{\prime \prime}_{2}\sin2\varphi_{2})^{2}+k\vert\beta^{\prime \prime}_{1,y}\sin\varphi_{1}\vert
\\
&+\frac{k}{2}(\beta^{\prime \prime}_{1,y}\sin\varphi_{1})^{2}+k\vert\beta^{\prime \prime}_{2,y}\sin\varphi_{2}\vert+\frac{k}{2}(\beta^{\prime \prime}_{2,y}\sin\varphi_{2})^{2}
\\
&+\chi k\vert\beta^{\prime \prime}_{1,y}\sin\varphi_{1}\vert \vert\beta^{\prime \prime}_{2,y}\sin\varphi_{2}\vert\bigg] dy,
\end{align*}
where $ \kappa = \frac{1}{2(1-\nu)} $. Solution of this minimization problem enables one to find $\beta ^{\prime \prime }_1(x,y)$ and $\beta ^{\prime \prime }_2(x,y)$ required for the reduction to 1-D theory. The analytical solution of the thickness problem can be found in the special case of symmetric slip systems with the angles $\varphi _1=\varphi \in (0,\pi /2)$, $\varphi_{2} = \pi - \varphi $. Obviously, $\beta^{\prime \prime}_{1} = \beta = -\beta^{\prime \prime}_{2} $ in this case due to the symmetry of the problem. Thus, functional $I_{3}$ simplifies to
\begin{equation}\label{3.6}
I_{3} = \int_{-h/2}^{h/2} [\frac{\kappa}{2}( -2v_{,xx}y+2\beta \sin 2\varphi)^{2}+2k\vert\beta_{,y} \sin\varphi\vert+k(1+\chi )(\beta_{,y}\sin\varphi)^{2} ]\, dy.
\end{equation}

Functional \eqref{3.6} is similar to the functional obtained in \citep{Le-nguyen2013}, so we present the solution of the thickness problem without detailed derivation. The minimizer $\beta(x,y)$ is
\begin{equation}\label{3.7}
\beta(x,y) = 
\begin{cases}
\beta_{m}(x)\quad &\text{for $y\in (-h/2,-l(x)/2)$},\\
\beta_{0}(x,y)\quad &\text{for $y\in (-l(x)/2,0)$},\\
-\beta(x,-y)\quad&\text{for $y\in (0,h/2)$}.
\end{cases}
\end{equation} 
Function $\beta_{0}(x,y)$ is found to be 
\begin{equation*}
\beta_{0}(x,y) = \frac{v_{,xx}}{\sin 2\varphi}\frac{1}{\eta}\left( \eta y - \frac{1}{\cosh \frac{\eta l}{2}}\sinh \eta y\right),
\end{equation*}
where
\begin{equation*}
\eta = 2\sqrt{\frac{2\kappa}{k(1+\chi)}}\cos \varphi .
\end{equation*}
Function $ \beta_{m}(x) $ is given by
\begin{equation*}
\beta_{m}(x) =  \frac{v_{,xx}}{\sin 2\varphi}\frac{1}{\eta}\left(-\frac{\eta l}{2} + \tanh \frac{\eta l}{2}\right) .
\end{equation*}
Finally, the length $l(x)$ should be found by solving the following transcendental equation
\begin{equation}\label{3.9}
\kappa v_{,xx}[ \frac{1}{4}(h^{2}-l^{2})+\frac{1}{\eta}( -\frac{\eta l}{2}+\tanh\frac{\eta l}{2})( h-l)]\sin 2\varphi -k\, \text{sign}(\beta_{0,y})\sin\varphi = 0. 
\end{equation}
If the length $l$ is small, then the sign of $v_{,xx}$ coincides with the sign of $\beta _{0,y}$ evaluated right from the point $y=-l/2$. For definiteness let $\beta _{0,y}(-l/2+0)>0$ so that $v_{,xx}>0$. Note that, if the curvature of the beam is constant, then $l(x)$ does not depend on $x$ and remains constant over the whole length of the beam. In this case \eqref{3.9} can be regarded as the equation for $v_{,xx}$ once $l$ is known. By integrating \eqref{3.5} in our special case and taking the constraints \eqref{3.4} into account, we obtain the displacements $u_x$ and $u_y$ in the form 
\begin{align*}
u_{x} &= -v_{,x}y,
\\
u_{y} &= v(x)+\frac{\gamma}{\gamma+2}\frac{1}{2}(y^{2}-\frac{h^{2}}{12})v_{,xx}+\frac{2}{\gamma+2}\left( \int^{y}_{0}\beta(x,\xi)\, d\xi-\Lambda \right) \sin2\varphi ,
\end{align*}
where $ \Lambda = \langle \int^{y}_{0}\beta(x,\xi)d\xi \rangle $.

Having found the solution of the thickness problem, let us now substitute the above formulas for the displacements together with the expression of $ \beta $ into the energy functional \eqref{3.2}. Keeping the asymptotically principal terms and integrating over the thickness we obtain the one-dimensional functional
\begin{align}\label{3.10}
J[v(x)] = \int^{L}_{0}\Phi(v_{,xx})\, dx - Mv_{,x}|_{x=L},
\end{align}
where the bending energy density reads
\begin{align*}
\Phi(\omega) = c_{1}\omega^{2} + c_{2}\omega.
\end{align*}
In these formulas we use $\omega $ to denote the curvature $v_{,xx}$, $M=\tau h^3/12$ is the resultant moment, and
\begin{align}\label{3.12}
&c_{1} = 2\int^{-l/2}_{-h/2}\frac{\kappa}{2}\left( 2q_{0}-2y\right)^{2}dy+2\int^{0}_{-l/2}\bigg[ \frac{\kappa}{2}\left( 2q-2y\right)^{2}+\frac{k(1+\chi)}{4\cos^{2}\varphi}(q_{,y})^{2}\bigg]dy, \nonumber
\\&c_{2} = 2\int^{0}_{-l/2}2kq_{,y}\frac{\sin\varphi}{\sin2\varphi}dy = -\frac{2k}{\cos\varphi}q_{0} ,
\end{align}
with
\begin{align}\label{3.13}
q(y) = \frac{1}{\eta}\left( \eta y - \frac{1}{\cosh \frac{\eta l}{2}}\sinh \eta y\right) ,
\end{align}
and $ q_{0} = q(-l/2) $. Note that, as the coefficients $c_1$ and $c_2$ depends on the curvature $\omega $ through $l$, the energy density is not quadratic with respect to $\omega $. Besides, if $q(y)=0$, then $c_1=\kappa h^3/6$ and $c_2=0$, so the obtained functional reduces to the classical 1-D functional of the elastic beam as expected \citep{Le1999}. 

Varying functional \eqref{3.10} with respect to the deflection $ v $, we obtain the differential equation of bending
\begin{align}\label{3.14}
m_{,xx} = 0,\quad m = \frac{\partial\Phi}{\partial\omega},
\end{align}
subject to the boundary conditions
\begin{align}\label{3.15}
\begin{cases}
&v(0) = 0,\quad v_{,x}(0) = 0,
\\&m(L)-M = 0,\quad m_{,x}(L) = 0.
\end{cases}
\end{align}
Equation \eqref{3.14}, together with the conditions \eqref{3.15}$_{2} $, implies that
\begin{align}\label{3.16}
m(\omega) = 2c_{1}\omega+\frac{dc_{1}}{d\omega}\omega^{2}+c_{2}+\frac{dc_{2}}{d\omega}\omega = M.
\end{align}
Since the bending moment $m$ is independent of $x$, the curvature must also be constant over the length of the beam. Consequently, $l$ and $\beta _0$ are independent of $x$ and the upper and lower layers of the beam are dislocation-free. Together with \eqref{3.9}, this equation determines the moment-curvature curve during the plastic bending. To plot this curve let us compute the derivatives of $c_1$ and $c_2$ with respect to $\omega $
\begin{align*}
\frac{dc_{1}}{d\omega} = \frac{dc_{1}}{dl}\frac{dl}{d\omega},\quad\frac{dc_{2}}{d\omega} = \frac{dc_{2}}{dl}\frac{dl}{d\omega}.
\end{align*}
From \eqref{3.9} we find that
\begin{align*}
\frac{d\omega}{dl} = \frac{4k\eta \tanh \tfrac{l\eta}{2} \left( \eta \left( h-l \right)\tanh \tfrac{l\eta}{2}+2 \right) }{\kappa \cos\varphi(h-l)^{2}\left( \eta(h-l)+4\tanh\tfrac{l\eta}{2}\right)^{2} }.
\end{align*}
This formula, together with \eqref{3.12}, enables one to determine $dc_1/d\omega $ and $dc_2/d\omega $, required for plotting the moment-curvature curve. 

The threshold value of curvature at which dislocations begin to nucleate is calculated by letting $ l $ go to zero in \eqref{3.9}. This yields
\begin{align*}
\omega _{en} = \dfrac{2k}{\kappa h^{2}\cos\varphi} .
\end{align*}
The threshold value of moment can be computed from \eqref{3.16}. Taking into account that, at $ l = 0$, $c_{1} = \kappa h^{3}/6$, $c_{2} = 0 $, while $ dc_{1}/d\omega=dc_{2}/d\omega=0 $, we get
\begin{align*}
M_{en} = \dfrac{2 kh}{3\cos\varphi}.
\end{align*}
Returning to the original physical quantities according to \eqref{3.1} we have
\begin{equation*}
M_{en}=\frac{\tau _{en}h^3a}{12}=\frac{2\mu kah}{3b\rho _s\cos \varphi },
\end{equation*}
which exhibits explicitly the size effect. Besides, $M_{en}$ goes to infinity as the shear modulus $\mu $ goes to infinity, so this coincides well with the result obtained by \citet{Ashby1970}. Thus, if $M<M_{en}$, then $\beta =0$, so no dislocation is nucleated and we have purely elastic solution. The plastic yielding begins at $M=M_{en}$.

Combining the elastic and plastic responses together, we get the moment-curvature relation in the following form
\begin{align*}
M=\begin{cases}
\tfrac{\kappa h^{3}}{3}\omega\quad &\text{for $M < M_{en}$},
\\2c_{1}\omega+\tfrac{dc_{1}}{d\omega}\omega^{2}+c_{2}+\tfrac{dc_{2}}{d\omega}\omega\quad &\text{for $M > M_{en}$}.
\end{cases}
\end{align*}
Knowing the curvature $\omega $ from $M$, we integrate the equation $v_{,xx}=\omega $ and use the boundary conditions \eqref{3.15}$_1$ to obtain the deflection of the beam 
\begin{align*}
v(x) = \dfrac{1}{2}\omega x^{2}.
\end{align*}

\section{Numerical simulations}

\begin{figure}[htb]
    \begin{center}
    \includegraphics[width=7cm]{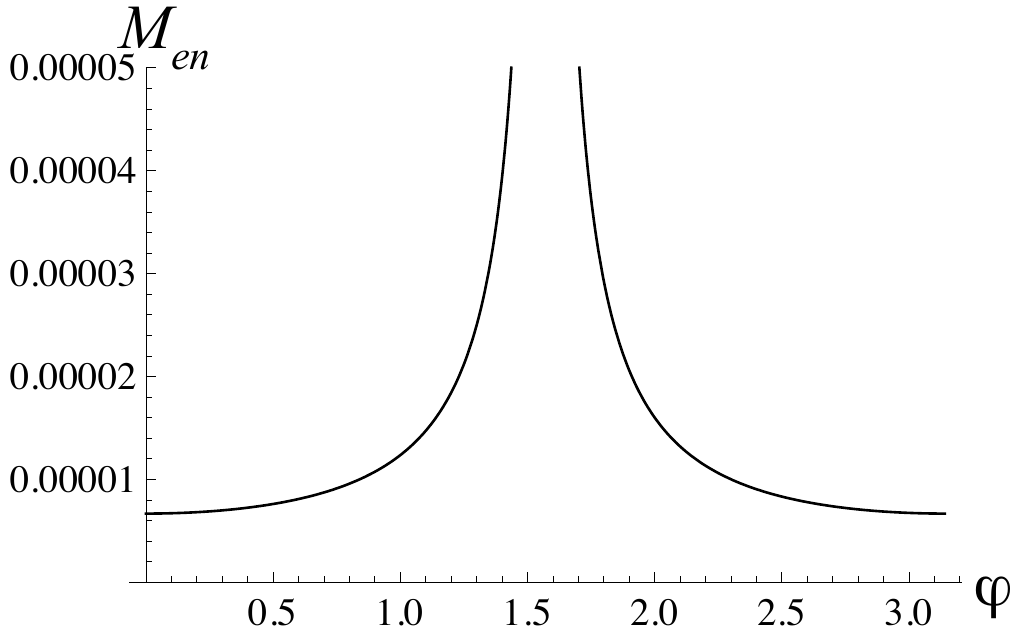}
    \end{center}
    \caption{Function $ M_{en}(\varphi) $}
    \label{fig:2}
\end{figure}

\begin{figure}[htb]
    \begin{center}
    \includegraphics[width=7cm]{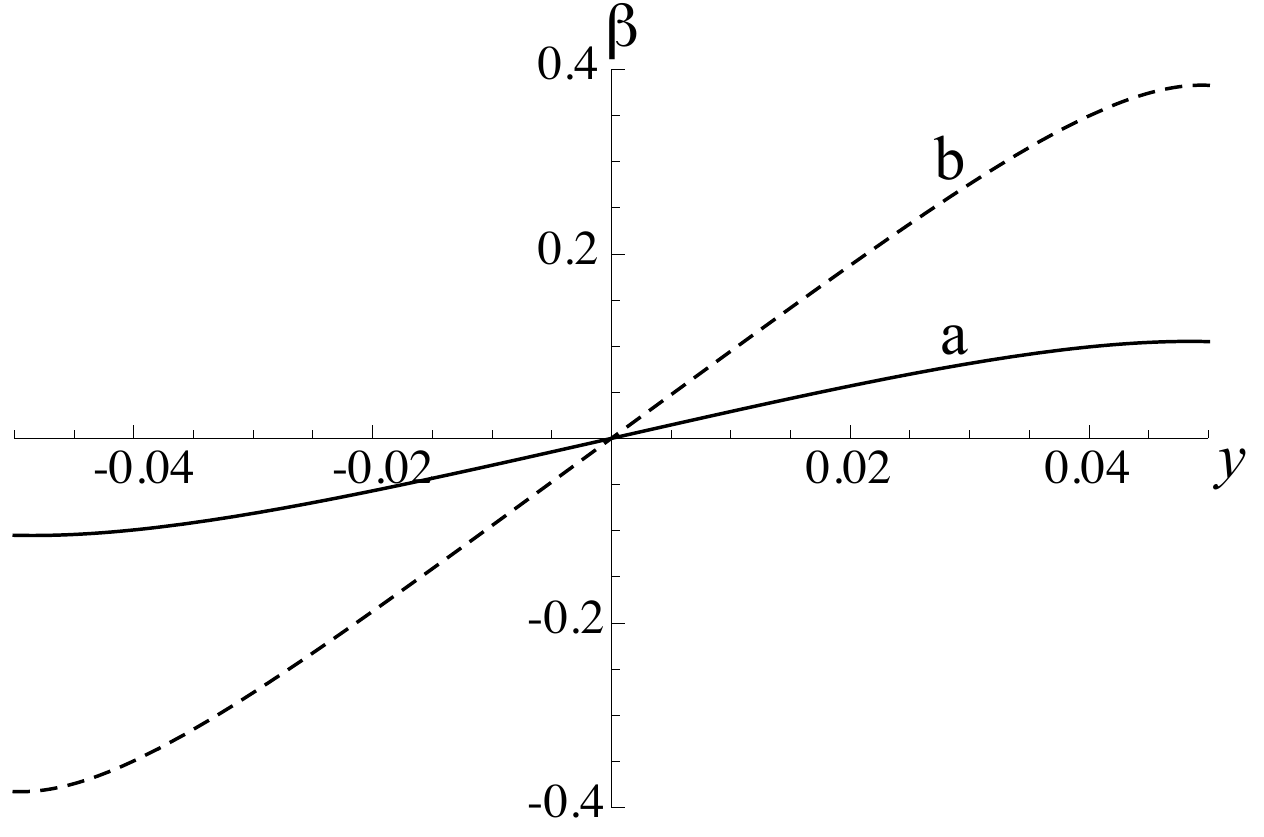}
    \end{center}
    \caption{Function $\beta(y)$ for $ M = 0.0001$ and (a) $\varphi = \pi/3$, (b) $\varphi = \pi/6$}
    \label{fig:3}
\end{figure}

In order to simulate the minimizer numerically, we choose $ h = 0.1$, $L = 10$, $\nu = 0.25$, $k = 10^{-4} $, and $\chi =0.8$. Fig.~\ref{fig:2} shows the plot of energetic threshold of the bending moment $M_{en}$ as function of the angle $\varphi $. It can be seen that the threshold moment becomes infinite as $\varphi $ goes to $\pi /2$ and has a local minimum at $\varphi =0$. It must be emphasized that this result is obtained under the assumption that the crystal admits two active symmetric slip systems. So, if $\varphi =\pi /2$ the slip systems merge into one that remains inactive during the whole process of bending because of its vanishing resolved shear stress (cf. with \citep{Le-nguyen2013}). However, if the slip systems are oriented unsymmetrically, then the one with largest Schmid factor will be activated at a certain finite bending moment.

For $M>M_{en}$ the plastic slip becomes non-zero. Fig.~\ref{fig:3} show the plots of plastic slip $\beta (y)$ for $M=0.0001$ at two different orientations of slip systems. The plastic strain vanishes on the middle line of the beam as expected and reaches its maximum and minimum at the traction-free faces. There are two small boundary layers near these faces where the plastic slip remains constant.

\begin{figure}[htb]
    \begin{center}
    \includegraphics[width=7cm]{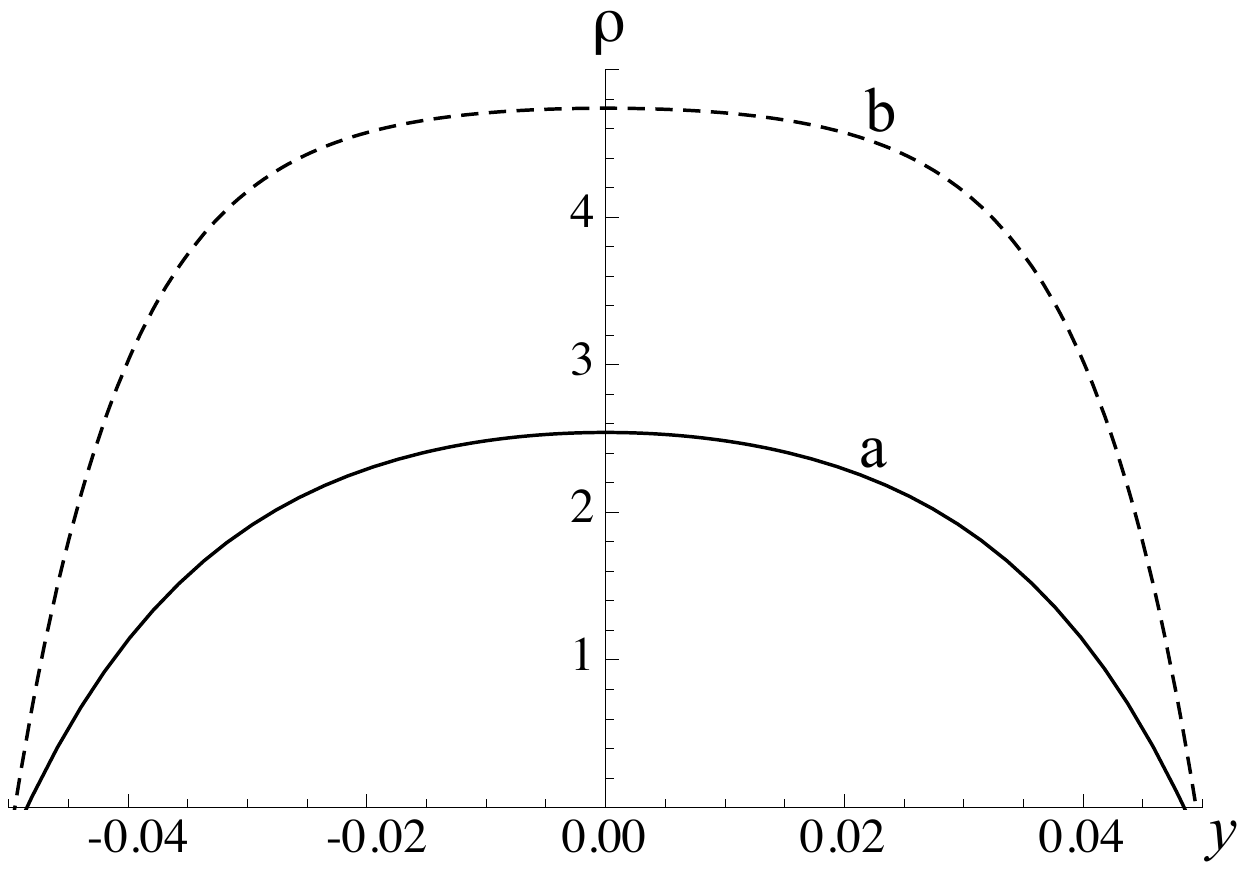}
    \end{center}
    \caption{Dislocation density for $ M = 0.0001$ and (a) $\varphi = \pi/3$ ($l\approx0.097$), (b) $\varphi = \pi/6$ ($l\approx0.099$)}
    \label{fig:4}
\end{figure}

\begin{figure}[htb]
    \begin{center}
    \includegraphics[width=7cm]{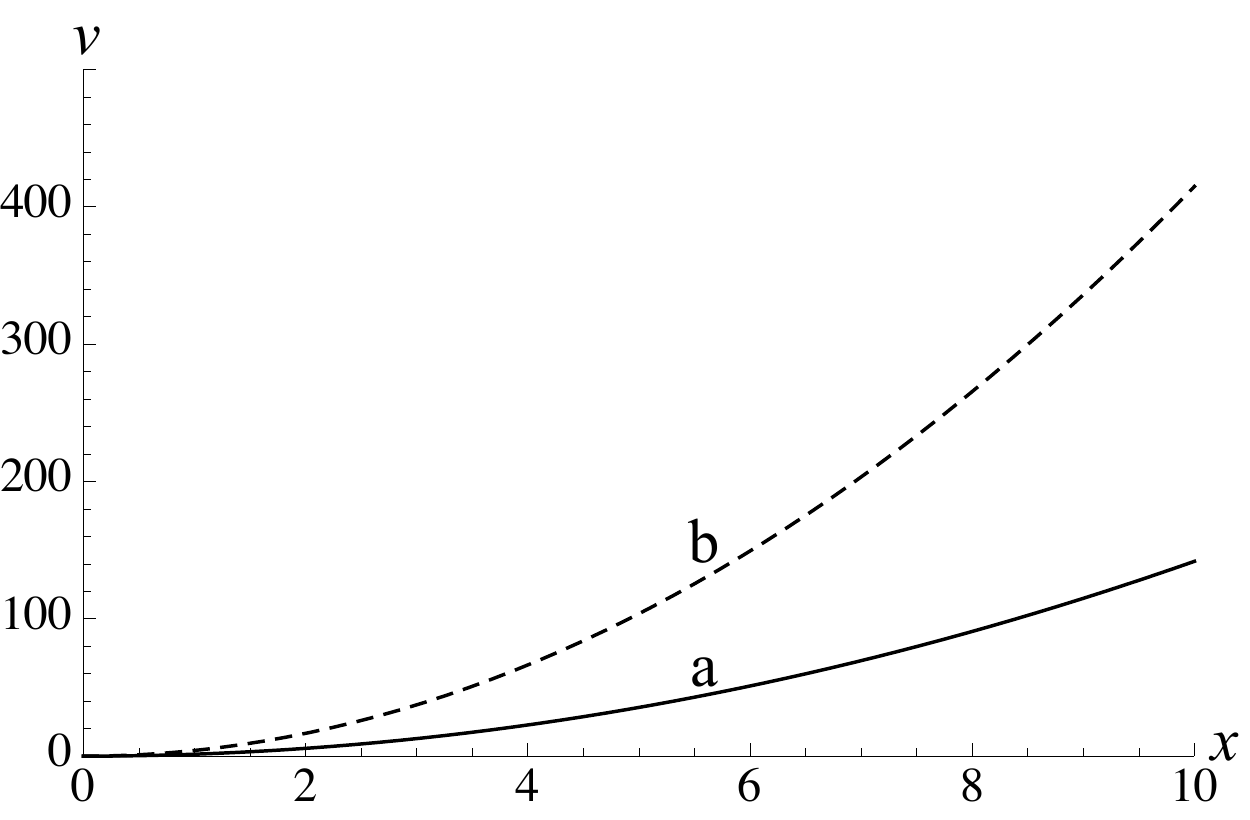}
    \end{center}
    \caption{Deflection of the beam}
    \label{fig:5}
\end{figure}

On Fig.~\ref{fig:4}, where the dimensionless dislocation density $\rho (y)=\beta _{,y}\sin \varphi $ is plotted for $M=0.0001$ and for two different angles, it is seen that the excess edge dislocations of the same sign of each group are concentrated in the middle of the beam thickness, with maximum dislocation density achieved at $y= 0$. Although no obstacle exists on the middle line, the repulsive forces between dislocations of the same sign prevent them from colliding. Thus, the high concentration of dislocations can be regarded as the dislocation pile-up against the middle line. The dislocation free zones are $y\in (-h/2,-l/2)$ and $y\in (l/2,h/2)$.

The dimensionless deflection of the beam, $v(x)$, is shown in Fig.~\ref{fig:5} for $M=0.0001$ and (a) $\varphi =\pi/3$, (b) $\varphi = \pi/6$. Since the curvature of the beam is constant, the thickness of the dislocation zone $l$ does not depend on $x$. 

\begin{figure}[htb]
    \begin{center}
    \includegraphics[width=7cm]{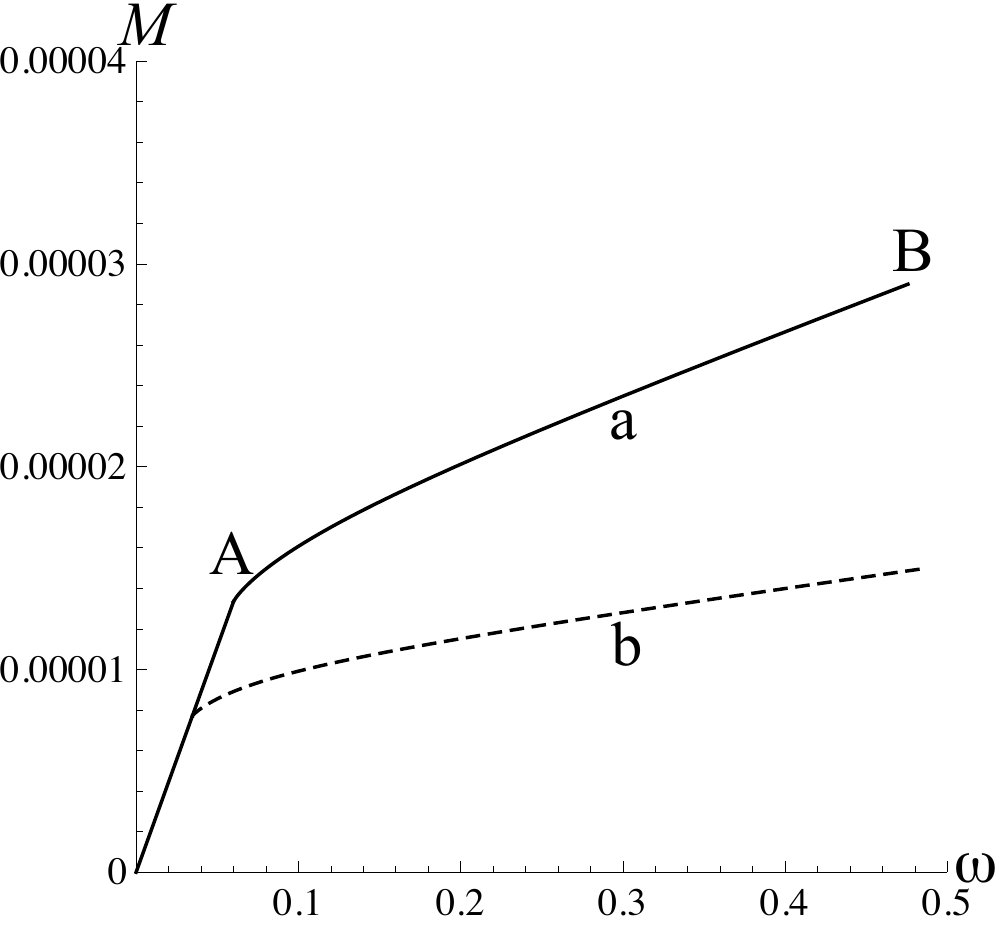}
    \end{center}
    \caption{Moment-curvature curve for: (a) $\varphi = \pi/3$, (b) $\varphi = \pi/6$}
    \label{fig:6}
\end{figure}

On Fig.~\ref{fig:6} we show the moment-curvature curve for $\varphi =\pi /3$ and $\varphi = \pi/6$. Up to the threshold moment $M_{en}$ (corresponding to point A on this figure) the moment-curvature curve is a straight line corresponding to the linear elastic beam theory. Then the curve becomes non-linear and increasing as $M$ increases and $l$ increases from zero to $h/2$. This nonlinear portion describes the work hardening due to the dislocation pile-up against the middle line of the beam. If the bending moment is increased from zero up to the moment $M_B$ corresponding to B (the loading case) we follow the moment-curvature curve from O through A to B. Now, if we unload the beam by decreasing the bending moment from $M_B$ to zero, the curvature and moment follow the same path BAO (in the inverse direction), and at the end of the unloading process no residual curvature of the beam is observed.

\section{Non-zero dissipation}

If the dissipation cannot be neglected, the problems reduce to minimizing the relaxed energy functionals \eqref{2.5}. The signs of the dissipation terms in this functional depend on whether $\dot{\beta }_\alpha >0$ or $\dot{\beta }_\alpha <0$, $\alpha =1,2$. However, for the beam bending it is easy to see that both cases occur simultaneously during the plastic deformations. Indeed, from the elementary beam theory (and also from the previous simulations) we know that, as the bending moment is increased (loading), $\dot{\beta }_1>0$ for $y>0$ and $\dot{\beta }_1<0$ for $y<0$, while $\dot{\beta }_2<0$ for $y>0$ and $\dot{\beta }_2>0$ for $y<0$. In contrary, if the bending moment is decreased (unloading or loading in the opposite direction), $\beta _1$ is either frozen or $\dot{\beta }_1>0$ for $y<0$ and $\dot{\beta }_1<0$ for $y>0$ while the opposite happens to $\beta _2$. Since these functionals differs from each other in the loading and unloading case, the case study must be done separately. 

The case study becomes simpler for the symmetric slip systems with $\varphi _1=\varphi \in (0,\pi /2)$, $\varphi _2=\pi -\varphi $. Due to the symmetry of the problem, we have $\beta _1=\beta =-\beta _2$. Therefore $K\, \text{sign}(\dot{\beta}_{1})\beta_{1} = K\, \text{sign}(\dot{\beta}_{2})\beta_{2}$. According to the above property, both terms can be replaced by $ K\, \text{sign} y \, \beta $. We use again the dimensionless quantities \eqref{3.1} and write the functional \eqref{2.5} in this special case in the form
\begin{align}
I_{d} = &\int^{L}_{0}\int^{h/2}_{-h/2} \bigg[ \dfrac{1}{2}\gamma\left(u_{x,x}+u_{y,y}\right) ^{2} +\left(u_{x,x}+\beta \sin2\varphi\right) ^{2}+\left(u_{y,y}-\beta \sin2\varphi\right) ^{2}\nonumber 
\\&+\dfrac{1}{2}\left( u_{x,y}+u_{y,x}\right)^{2}+k \vert \beta_{,x}\cos\varphi+\beta_{,y}\sin\varphi\vert+\dfrac{1}{2}k\left( \beta_{,x}\cos\varphi+\beta_{,y}\sin\varphi\right)^2 \nonumber
\\& +k \vert \beta_{,x} \cos \varphi - \beta_{,y}\sin\varphi\vert+\dfrac{1}{2}k\left( \beta_{,x}\cos\varphi-\beta_{,y}\sin\varphi\right)^{2} \nonumber
\\&+k\chi \vert \beta_{,x}\cos \varphi+\beta_{,y}\sin \varphi \vert \cdot \vert \beta_{,x}\cos\varphi-\beta_{,y}\sin \varphi \vert +2\epsilon \, \text{sign}y\, \beta \bigg]dxdy \nonumber
\\&+\int^{h/2}_{-h/2}\tau yu_{x}|{x=L}dy, \label{5.0}
\end{align}
where $ \epsilon = K/\mu $.  As compared to the previous functional \eqref{3.2} the additional term $2\epsilon \, \text{sign}\, y\, \beta $ does not belong to the asymptotically principal terms. Therefore, up to the second step of the variational-asymptotic procedure this term does not have any influence on the inner asymptotic expansion. At the third step, we are looking for the minimize in the form \eqref{3.3} such that the constraints \eqref{3.4} are fulfilled. Fixing $ \beta^{\prime \prime } $ and minimizing the relaxed energy with respect to $ u^{\prime \prime }_{x} $ and $ u^{\prime \prime }_{y} $ we find them in the form \eqref{3.5}. Then, the functional reduces to
\begin{align*}
I_{d} = &\int_{-h/2}^{h/2} \bigg[ \dfrac{\kappa}{2} \left( -2v_{,xx}y+2\beta^{\prime \prime } \sin 2\varphi \right)^{2}+2k\vert \beta^{\prime \prime }_{,y} \sin \varphi \vert +k(\beta^{\prime \prime }_{,y}\sin \varphi)^{2}\nonumber
\\&+\chi k\left( \beta^{\prime \prime }_{,y}\sin \varphi \right)^{2}+2\epsilon \, \text{sign} y \, \beta^{\prime \prime } \bigg] dy
\end{align*}
Now the term $ 2\epsilon\, \text{sign}y \, \beta^{\prime \prime } $ should be kept because it has the same order as the cross term $ -4\kappa v_{,xx}y\beta^{\prime \prime }\sin 2\varphi $. Up to an unessential  constant we may rewrite this functional as
\begin{align}\label{5.1}
I_{d} = &\int_{-h/2}^{h/2} \bigg[ \dfrac{\kappa}{2}\left( -2v_{,xx}y+\dfrac{\epsilon \, \text{sign}y }{\kappa \sin 2\varphi}+2\beta^{\prime \prime } \sin 2\varphi\right)^{2}+2k\vert\beta^{\prime \prime }_{,y} \sin\varphi\vert \nonumber
\\&+k(1+\chi )(\beta^{\prime \prime }_{,y}\sin\varphi)^{2}\bigg]dy.
\end{align}

Functional \eqref{5.1} is similar to the functional obtained in \citep{Le-nguyen2013} for the case of non-zero dissipation, so we again present the solution of the thickness problem without detailed derivation. We use the same Ansatz \eqref{3.7} for $\beta $ in this case. Then for function $\beta _0(y)$ we find 
\begin{align}\label{5.2}
\beta_{0}(y)&=\dfrac{v_{,xx}}{\sin 2\varphi}\dfrac{1}{\eta}\left(\eta y-\dfrac{1}{\cosh\tfrac{\eta l}{2}}\sinh\eta y \right) \nonumber
\\
&+\dfrac{\epsilon}{2\kappa \sin^{2}2\varphi}\left(1-\cosh\eta y-\tanh\tfrac{\eta l}{2}\sinh\eta y \right) .
\end{align}
Computing $ \beta_{0}(y) $ at $ y = -l/2 $ we get
\begin{align*}
\beta_{m}=\dfrac{v_{,xx}}{\sin 2\varphi\eta}\left( -\dfrac{\eta l}{2}+\tanh \dfrac{\eta l}{2}\right)+\dfrac{\epsilon}{2\kappa \sin^{2}2\varphi}\left( 1-\dfrac{1}{\cosh\tfrac{\eta l}{2}}\right)  
\end{align*}
The transcendental equation for $l$ reads
\begin{align*}
&\kappa v_{,xx} [ \dfrac{1}{4}(h^{2}-l^{2})+\dfrac{1}{\eta}( -\dfrac{\eta l}{2}+\tanh\dfrac{\eta l}{2})(h-l)] \sin 2\varphi \nonumber
\\&-\dfrac{\epsilon}{2}\dfrac{1}{\cosh\tfrac{\eta l}{2}}(h-l)-k \, \text{sign}(\beta_{1,y})\sin\varphi = 0.
\end{align*}

Knowing $\beta $, we can now determine the displacement field in accordance with \eqref{3.5}. Then, substituting this field together with $\beta (y)$ from \eqref{3.7} and \eqref{5.2} into the energy functional \eqref{5.0}, keeping the asymptotically principal terms and integrating over the thickness we obtain the 1-D functional 
\begin{align}\label{5.4}
J_d[v(x)] = \int^{L}_{0}\Phi(v_{,xx})dx - Mv_{,x}|_{x=L},
\end{align}
with the bending energy density
\begin{align*}
\Phi(\omega) = c_{1}\omega^{2} + c_{2}\omega.
\end{align*}
The coefficient $ c_{1} $ and $ c_{2} $ are given by
\begin{align*}
c_{1} &= 2\int^{-l/2}_{-h/2} \dfrac{\kappa}{2}\left( 2q_{0}-2y\right)^{2}dy+2\int^{0}_{-l/2} \bigg[\dfrac{\kappa}{2}\left( 2q-2y\right)^{2}+\dfrac{k(1+\chi)}{4\cos^{2}\varphi}(q_{,y})^{2}\bigg]dy,
\\
c_{2} &= 2\int^{0}_{-l/2} \bigg[ \dfrac{k}{\cos\varphi}q_{,y}+\dfrac{\epsilon}{\kappa \sin2\varphi}(\dfrac{k(1+\chi)}{4\cos^{2}\varphi}q_{,y}p_{,y}+\kappa(p(y)-1)(2q(y)
\\
&-2y) )\bigg] dy+2\int^{-l/2}_{-h/2}\bigg[\dfrac{\epsilon}{\sin2\varphi}(p_{0}-1)(2q_{0}-2y)\bigg] dy ,
\end{align*}
with 
\begin{align*}
q(y)=\dfrac{1}{\eta}(\eta y-\dfrac{\sinh\eta y }{\cosh\tfrac{\eta l}{2}}),\quad p(y)=1-\cosh\eta y-\tanh\dfrac{\eta l}{2}\sinh\eta y ,
\end{align*}
and $ p_{0}=p(-l/2),\,q_{0}=q(-l/2) $. Note that the coefficients $c_1$ and $c_2$ depends on the curvature $\omega $, so the energy density is not quadratic with respect to $\omega $.

Varying functional \eqref{5.4} with respect to the deflection $v(x)$ we get the same equilibrium equation and boundary conditions as in Section 4. Consequently, the moment-curvature relation \eqref{3.16} remains valid if $c_1$ and $c_2$ from above are substituted. It turns out that the straight line $M=\kappa h^3\omega /3$ corresponding to the purely elastic solution does not intersect the moment-curvature curve corresponding to the solution with dislocations at $l=0$ for the case with non-zero dissipation. The threshold value for the curvature must therefore be calculated by solving the system of equations
\begin{equation}\label{5.5a}
\omega (l_d)=\omega_{d}, \quad m(l_d)=\kappa h^3\omega _d/3,
\end{equation}
where $l_d$ and $\omega _d$ are regarded as unknowns.\footnote{Formula for $\omega _d$ and Figure 7 plotting $M_d$ in \citep{Le-nguyen2013} are not correct and should be amended in accordance with \eqref{5.5a}.} Using \eqref{3.16} we determine also the dissipative threshold value for the bending moment $M_d$. Computing the curvature from the given moment, we find the deflection by integrating the equation $v_{,xx}=\omega $.

Consider now the following loading/unloading process. We first bend the beam slowly and successively by increasing the moment until some maximum value $M_*>M_d$ such that at the end of the loading process the plastic slip becomes $\beta _*$. Then we unload the beam by reducing the bending moment back to zero. Assuming that dislocations are frozen during this unloading process with $\beta =\beta _*$ and $\dot{\beta }=0$, we have to find the displacements of the beam by minimizing the energy functional
\begin{multline}\label{5.6}
I_{d} = \int^{L}_{0}\int^{h/2}_{-h/2} \bigg[ \dfrac{1}{2}\gamma\left(u_{x,x}+u_{y,y}\right) ^{2} +\left(u_{x,x}+\beta \sin 2\varphi\right) ^{2}+\left(u_{y,y}-\beta \sin 2\varphi\right) ^{2} 
\\
+\dfrac{1}{2}\left( u_{x,y}+u_{y,x}\right)^{2}+k\vert\beta_{,x} \cos\varphi+\beta_{,y}\sin\varphi\vert+\dfrac{1}{2}k\left( \beta_{,x}\cos\varphi+\beta_{,y}\sin\varphi\right)
\\
+k\vert\beta_{,x}\cos\varphi-\beta_{,y}\sin\varphi\vert+\dfrac{1}{2}k\left( \beta_{,x}\cos\varphi-\beta_{,y}\sin\varphi\right)^{2}+k\chi\vert\beta_{,x}\cos\varphi 
\\
+\beta_{,y}\sin\varphi\vert \cdot\vert\beta_{,x}\cos\varphi-\beta_{,y}\sin\varphi\vert +2\epsilon \, \text{sign}y\, \beta\bigg] dxdy+\int^{h/2}_{-h/2}\tau yu_{x}|_{x=L}dy.
\end{multline}
where, now, $\beta _*$ is fixed and does not subject to the variation. Since functional \eqref{5.6} is functional \eqref{3.2} with $\varphi_{1}=\pi-\varphi_{2}=\varphi $ and $ \beta_{1}=-\beta_{2} =\beta_{*}$, exactly the same variational-asymptotic analysis as in Sections 3 can be repeated leading to the following displacement field
\begin{align*}
&u_{x} = -v_{,x}y,
\\&u_{y} = v(x)+\tfrac{\gamma}{\gamma+2}\tfrac{1}{2}(y^{2}-\tfrac{h^{2}}{12})v_{,xx}+\tfrac{2}{\gamma+2}\left( \int^{y}_{0}\beta _*(x,\xi)d\xi-\Lambda _* \right)\sin2\varphi ,
\end{align*}
where  $ \Lambda _*= \langle \int^{y}_{0} \beta _*(x,\xi)d\xi \rangle$. We substitute this displacement field into functional \eqref{5.6} and keep the asymptotically principal terms. Integrating over the thickness and neglecting the terms containing the known function $ \beta_{*} $ only, we get 1-D functional \eqref{5.4} with different coefficients $ c_{1} $ and $ c_{2} $
\begin{align*}
c_{1}=\kappa h^{3}/6,\quad c_{2}=2\int^{0}_{-l/2}\kappa\omega_{*}q_{*}(y)(-4y)dy+2\int^{-l/2}_{-h/2}\kappa\omega_{*}q_{0*}(y)(-4y)dy,
\end{align*}
where $ \omega_{*} $ is the curvature corresponding to $ M_{*} $ and
\begin{align*}
q_{*}(y)=\dfrac{1}{\eta}\left( \eta y-\dfrac{1}{\cosh\tfrac{\eta l_{*}}{2}}\sinh\eta y\right),\,q_{0*}=q_{*}(-l_*/2),
\end{align*}
$l_*$ being the length corresponding to $\beta _*$. 
Thus, during this unloading process the moment-curvature relation takes the form
\begin{align*}
m = 2c_{1}\omega+c_{2}=M,
\end{align*}
so that, at the end of the unloading when $M=0$, the residual curvature is
\begin{align*}
\omega_{1}=-\dfrac{c_{2}}{c_{1}}.
\end{align*}
Note that, although the bending moment after unloading is zero, the elastic strain is not, so there is still some eigenstress in the beam. 

\section{Numerical simulation}

In order to simulate the minimizer numerically, we choose the same numerical values for $h$, $L$, $\nu $, $k$, $\chi $ as in Section 4. For $\epsilon =K/\mu $ we choose the value $\epsilon =0.0001$. In order to find the dissipative threshold for the bending moment we need to solve equations \eqref{5.5a} with respect to $\omega _d$ and $l_d$. Then $M_d=\kappa h^3\omega _d/3$. Numerical simulations give for instance
\begin{equation*}
\omega _d=0.0413739, \quad l_d=0.0192525 \quad \Rightarrow M_d= 0.00001488 \quad \text{for $\varphi =\pi /3$},
\end{equation*}
and 
\begin{equation*}
\omega _d=0.0669535, \quad l_d=0.0198977 \quad \Rightarrow M_d= 0.91942\times 10^{-5} \quad \text{for $\varphi =\pi /6$}.
\end{equation*}
Thus, if we increase $M$ steadily, then at the onset of plastic yielding the plastic slip and the total number of dislocations turn out to be finite. This could be explained physically in the following way: in the presence of non-zero dissipation dislocations should accumulate enough to give the crystal lower ``relaxed'' energy as compared to the energy of purely elastic bending. This means also that the states with smaller number of dislocations present energy barriers to be overcome, and to have the smaller ``relaxed'' energy the crystal should jump to the certain finite dislocation density.  

\begin{figure}[htb]
    \begin{center}
    \includegraphics[width=7cm]{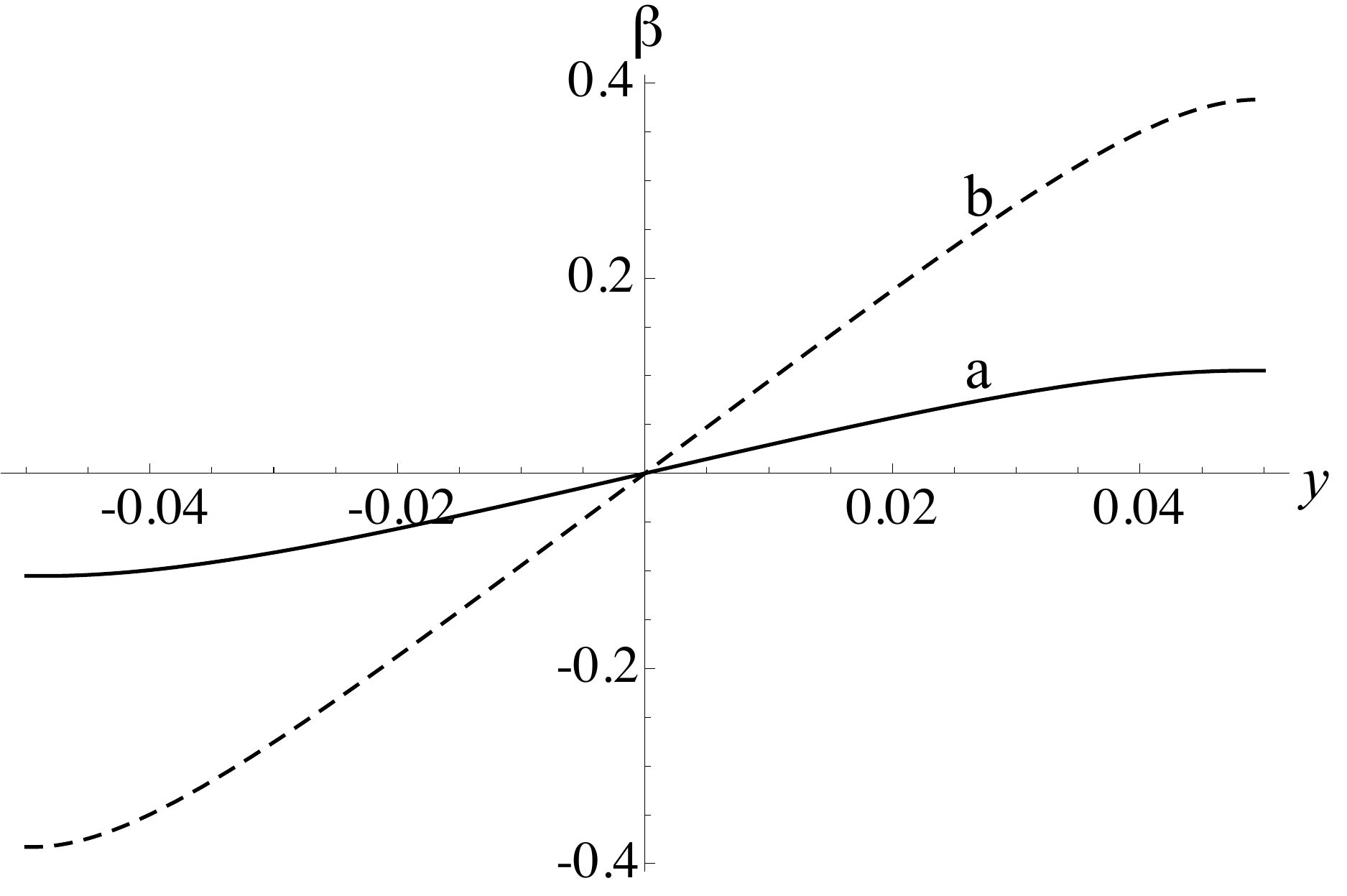}
    \end{center}
    \caption{Function $\beta(y)$ for $ M = 0.0001$ and (a) $\varphi = \pi/3$, (b) $\varphi = \pi/6$}
    \label{fig:8}
\end{figure}

For $M>M_{d}$ the plastic slip becomes non-zero. Fig.~\ref{fig:8} shows the plots of plastic slip $\beta (y)$ for $M=0.0001$ at two different orientations of slip systems. The plastic strain is zero on the middle line of the beam and reaches its maximum and minimum at the free faces. There are two small boundary layers near these faces where the plastic slip remains constant.

\begin{figure}[htb]
    \begin{center}
    \includegraphics[width=7cm]{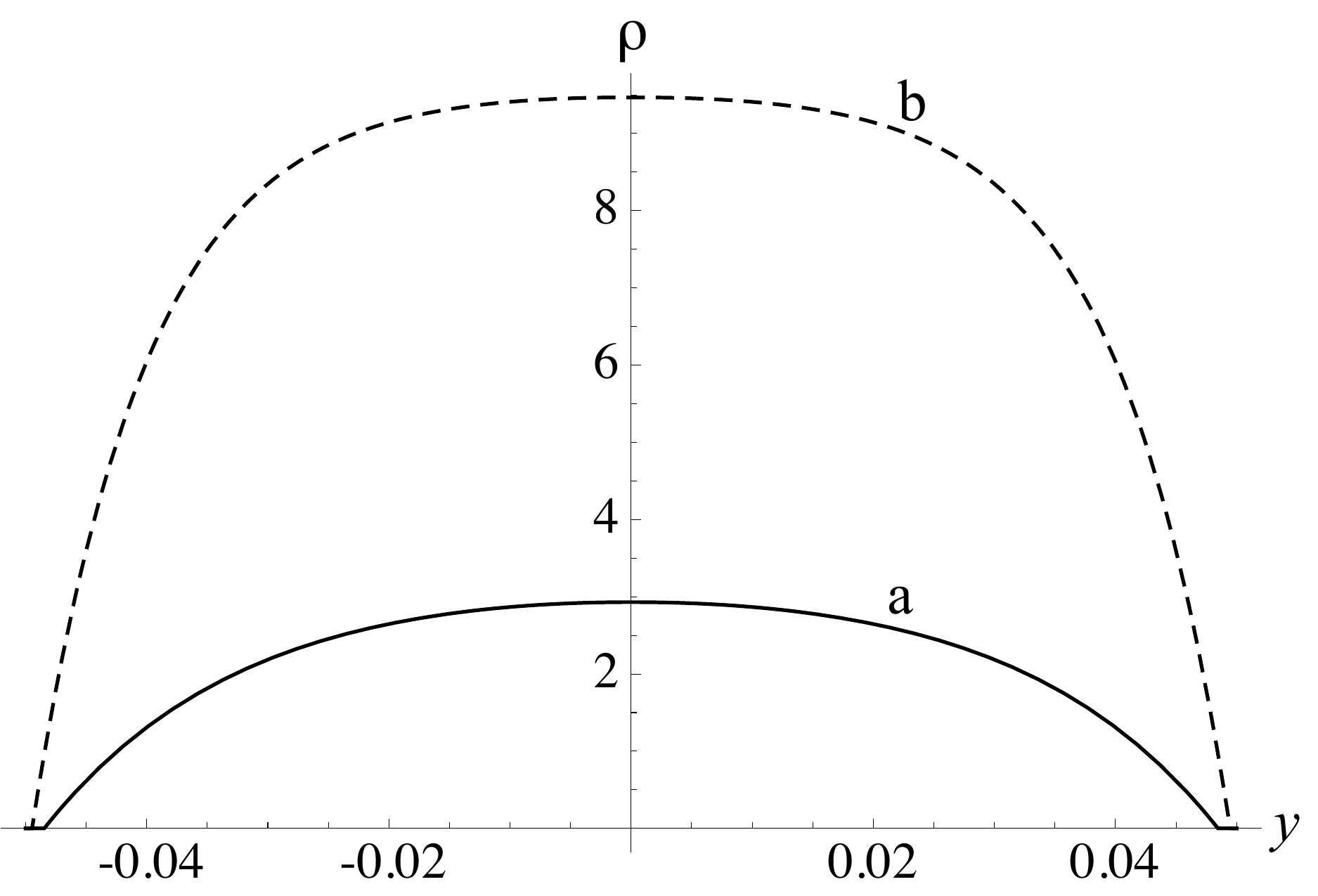}
    \end{center}
    \caption{Dislocation density for $ M = 0.0001$ and (a) $\varphi = \pi/3$ ($l\approx0.097$), (b) $\varphi = \pi/6$ ($l\approx0.099$)}
    \label{fig:9}
\end{figure}

On Fig.~\ref{fig:4}, where the dimensionless dislocation density $\rho (y)=\beta _{,y}\sin \varphi $ is plotted for $M=0.0001$ and for two different angles, it is seen that the excess edge dislocations of the same sign of each group are concentrated in the middle of the beam thickness, with maximum dislocation density achieved at $y= 0$. Although no obstacle exists on the middle line, the repulsive forces between dislocations of the same sign prevent them from colliding. Thus, the high concentration of dislocations can be regarded as the dislocation pile-up against the middle line. The dislocation free zones are $y\in (-h/2,-l/2)$ and $y\in (l/2,h/2)$.

\begin{figure}[htb]
    \begin{center}
    \includegraphics[width=7cm]{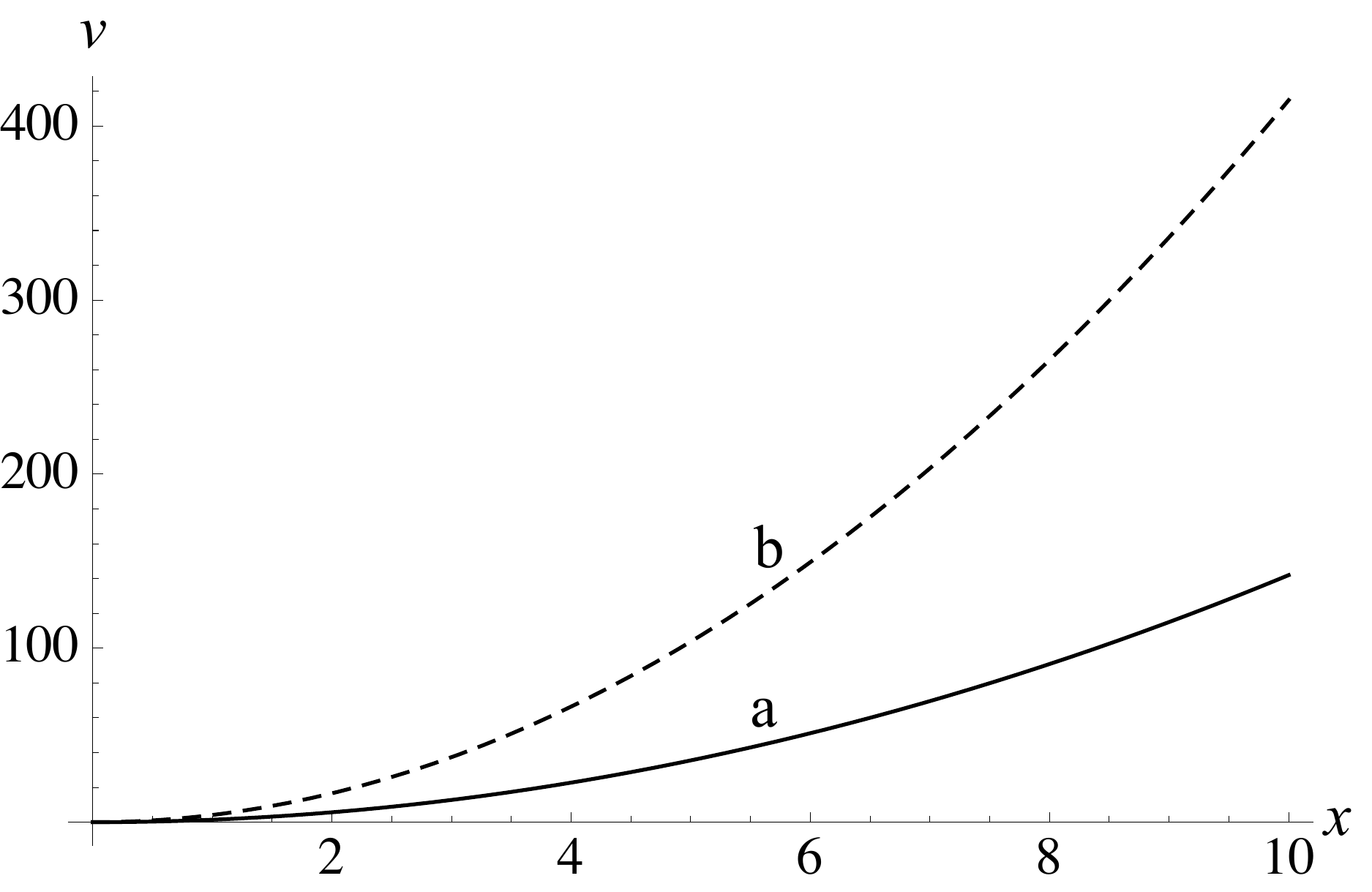}
    \end{center}
    \caption{Deflection of the beam}
    \label{fig:10}
\end{figure}

\begin{figure}[htb]
    \begin{center}
    \includegraphics[width=7cm]{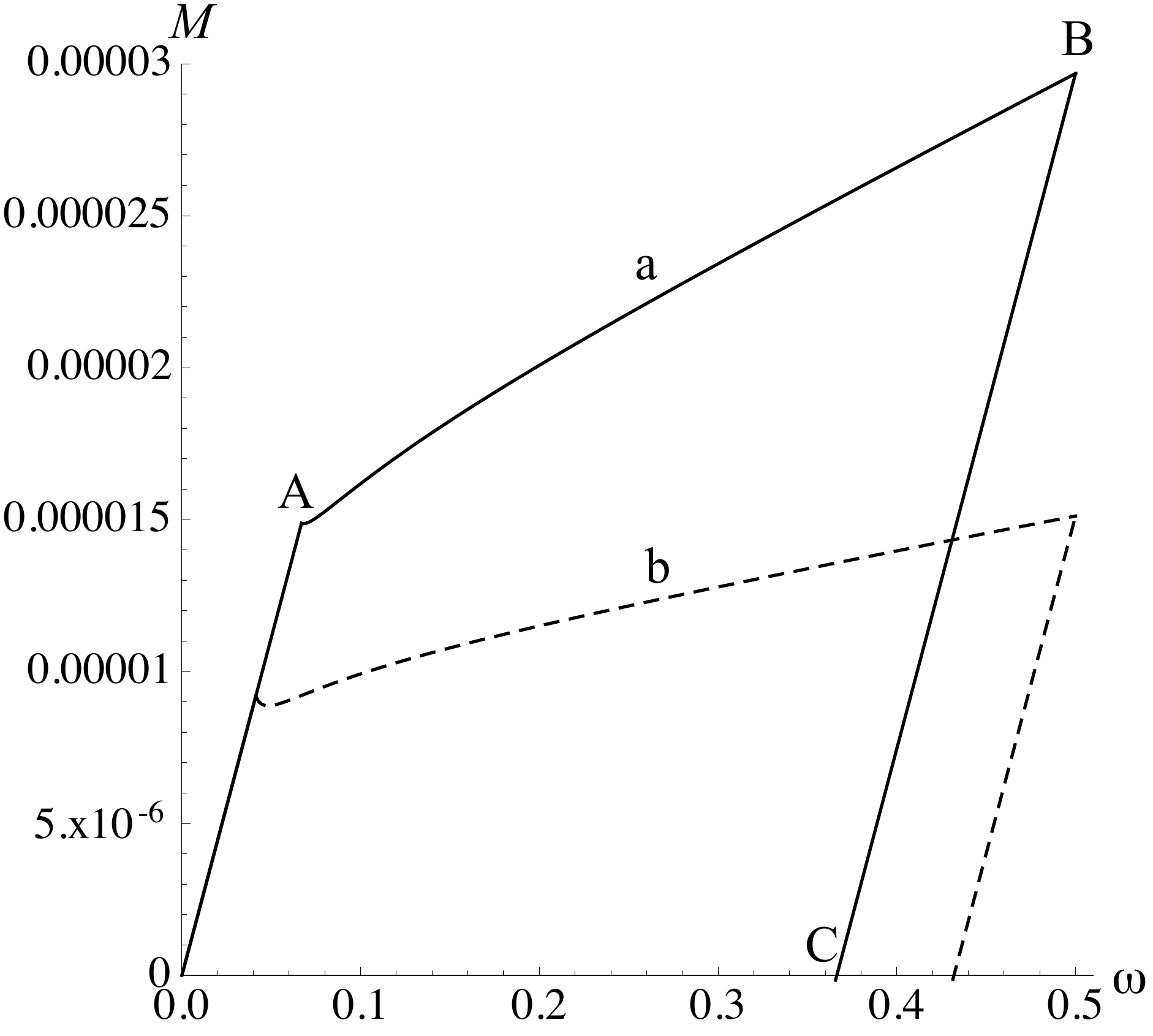}
    \end{center}
    \caption{Moment-curvature curve during loading and unloading for: a) $\varphi = \pi/3$, b) $\varphi = \pi/6$.}
    \label{fig:11}
\end{figure}

The deflection of the beam $ v(x) $ is shown in Fig.~\ref{fig:10} for  $ M = 0.0001$ and (a) $\varphi = \pi/3,$ (b) $\varphi = \pi/6$. Since the curvature of the beam is constant, the thickness of the dislocation zone $l$ does not depend on $x$.

On Fig.~\ref{fig:11} we show the moment-curvature curve for $\varphi =\pi /3$ and $\varphi = \pi/6$, respectively. Up to the threshold moment $M_{d}$ (corresponding to point A on this figure) the moment-curvature curve is a straight line corresponding to the linear elastic beam theory. Then the curve becomes non-linear and, in contrast to the zero dissipation case, also non-monotone as $\omega $ increases as can be seen in curve (b). At the onset of the plastic yielding we can observe a small softening behavior of the moment-curvature curve. The further monotonously increasing portion of the moment-curvature curve describes the work hardening due to the dislocation pile-up against the middle line of the beam. If the bending moment is increased from zero up to the moment $M_B$ corresponding to B (the loading case) we follow the moment-curvature curve from O through A to B (with a possible jump in $\omega $ to pass the softening portion). If the beam is then unloaded, the moment-curvature curve becomes a straight line BC with the same slope like that of OA as shown in Fig.~\ref{fig:11}.  

\section{Discussion and outlook}
In this paper the one-dimensional theory of bending of a single crystal beam with two active slip systems has been developed. The one-dimensional energy of bending can be found after the solution of the thickness problem. The latter has been solved for the symmetric slip systems. The threshold bending moment exhibiting the size effect has been found for the case without and with dissipation. We have found also the dislocation density and the moment-curvature curves at loading and unloading. It would be interesting to verify the obtained results with the measurement of excess dislocation density in a bent beam by the method of orientation imaging microscopy developed recently in \citep{Kysar2010}. The specimen could be for instance the fcc single crystal beam having the axis parallel to $[\bar{1}10]$-direction and bent about the $[110]$ direction. For the unsymmetrically oriented slip systems the thickness problem can only be solved numerically. This will be addressed in our forthcomming paper.

\bigskip
\noindent {\it Acknowledgments}

The financial support by the German Science Foundation (DFG) through the research project LE 1216/4-2 is gratefully acknowledged.

\end{document}